\documentclass[aps,prd,twocolumn,showpacs,superscriptaddress,nofootinbib]{revtex4-1}  
\usepackage{graphicx}  
\usepackage{dcolumn}   
\usepackage{amssymb,amsmath}   
\usepackage{color}

\newcommand{\xvect}{{\bf x}_{\bot}}
\newcommand{\pvect}{{\bf p}_{\bot}}
\newcommand{\kvect}{{\bf k}_{\bot}}

\newcommand{\ihat}{\hat{\imath}}

\newcommand{\dInt}{\text{d}}

\newcommand{\OneOverQ}{Q^{-1}}

\begin{document}

\title{Turbulent thermalization process in heavy-ion collisions at ultrarelativistic energies}

\author{J.~Berges}
\affiliation{Institut f\"{u}r Theoretische Physik, Universit\"{a}t Heidelberg, Philosophenweg 16, 69120 Heidelberg, Germany}
\affiliation{ExtreMe Matter Institute (EMMI), 
GSI Helmholtzzentrum f\"ur Schwerionenforschung GmbH, 
Planckstra\ss e~1, 64291~Darmstadt, Germany}

\author{K.~Boguslavski}
\affiliation{Institut f\"{u}r Theoretische Physik, Universit\"{a}t Heidelberg, Philosophenweg 16, 69120 Heidelberg, Germany}

\author{S.~Schlichting}
\email{soeren@kaiden.de}
\affiliation{Brookhaven National Laboratory, Physics Department, Bldg. 510A, Upton, NY 11973, USA}

\author{R.~Venugopalan}
\affiliation{Brookhaven National Laboratory, Physics Department, Bldg. 510A, Upton, NY 11973, USA}

\begin{abstract}
The non-equilibrium evolution of heavy-ion collisions is studied in the limit of weak coupling at very high energy employing lattice simulations of the classical Yang-Mills equations. Performing the largest 
classical-statistical simulations to date, we find that the dynamics of the longitudinally expanding plasma becomes independent of the details of the initial conditions. After a transient regime dominated by plasma instabilities and free streaming, the subsequent space-time evolution is governed by a nonthermal fixed point, where the system exhibits the self-similar dynamics characteristic of wave turbulence. This allows us to distinguish between different kinetic scenarios in the classical regime. Within the accuracy of our simulations, the scaling behavior found is consistent with the ``bottom-up" thermalization scenario~\cite{bib:BMSS}.

\end{abstract}

\maketitle

\textit{Introduction.} The non-equilibrium dynamics of the quark-gluon plasma, and how it proceeds to thermalization, is an outstanding problem in quantum chromodynamics (QCD). This is partly because the coupling constant at realistic collider energies may not be small enough for reliable weak coupling computations, but also because of the complex interplay of the different scales involved in the thermalization process. 

Significant progress has been achieved in our understanding of expanding non-Abelian plasmas in the two limiting cases of very strong coupling and very weak coupling. The former has been studied in supersymmetric Yang-Mills theories employing the gauge-string duality. The results in this case indicate the important role of anisotropies in the longitudinally expanding system even at the transition to the hydrodynamic regime~\cite{Heller:2011ju}. Ab initio computations are also feasible in QCD in the weak-coupling limit $\alpha_s \ll 1$. In this paper, we will address the early stages of the thermalization process in an expanding non-Abelian plasma in the weak coupling limit.

The colliding nuclei in weak coupling asymptotics are described in the Color Glass Condensate (CGC) framework~\cite{bib:CGC}. In such a collision, a non-equilibrium Glasma~\cite{bib:Glasma} of highly occupied gluon fields with typical momentum $Q$ is formed immediately after the collision. Since the characteristic occupancies $\sim 1/\alpha_s (Q)$ are large, the gauge fields are strongly correlated even for small gauge coupling. In such highly occupied systems, dynamical quantum effects are suppressed at early times and the nonequilibrium quantum dynamics can be accurately mapped onto a classical-statistical problem. The latter can be rigorously solved using real-time lattice simulation techniques. 

This real-time computation in lattice gauge theory is a formidable task. The work reported on here represents by far the largest numerical effort in this respect to date. Many of the details of the computation discussed here are given in a longer companion paper~\cite{Berges:2013fga} which considers non-Abelian plasmas in both expanding and non-expanding geometries. In this paper, we restrict our attention to expanding non-Abelian plasmas\footnote{The expansion of the plasma in the longitudinal direction $x^3$ at time $x^0$ is conveniently discussed in terms of the proper time $\tau \equiv \sqrt{(x^0)^2-(x^3)^2}$ and rapidity $\eta \equiv \text{atanh}(x^3/x^0)$. We denote the transverse coordinates by $\xvect = (x^1,x^2)$.} that are relevant to the collision of nuclei at very high energies. Our focus will be on the key results from the computations and the lessons one may draw with regard to the thermalization process at weak coupling. 

While much recent work has been devoted to the ab initio calculation of initial conditions in heavy ion collisions at proper times $\tau \lesssim 1/Q$ in the CGC framework~\cite{Dusling:2011rz,Gale:2012rq,Epelbaum:2013waa}, our focus here will be to understand the dependence of the evolution on the initial conditions at times $\tau \gtrsim 1/Q$. The properties of the system at these later times are markedly different from the initial conditions at earlier times. This is because the early time behavior of the Glasma is governed by plasma instabilities as demonstrated by a large number of studies in the literature~\cite{Mrowczynski:1993qm,Mrowczynski:2005ki,Romatschke:2003ms,Romatschke:2004jh,Arnold:2003rq,Arnold:2004ti,Attems:2012js}. 

In the background of the highly anisotropic Glasma fields, plasma instabilities lead to a rapid growth of vacuum fluctuations of the initial state~\cite{Romatschke:2005pm,Romatschke:2006nk,Romatschke:2005ag,Fukushima:2011nq,Berges:2012cj}. Even though these vacuum fluctuations are `quantum' in origin, their dynamics is classical-statistical in nature and can be accurately described in terms of Gaussian fluctuations around the CGC determined background fields~\cite{Dusling:2011rz,Epelbaum:2013waa}. In each configuration of this classical-statistical ensemble vacuum fluctuations are generated by random seeds. While at initial times $\tau \lesssim Q^{-1}$ these vacuum fluctuations are small, they become on the order of the classical background fields at times $\tau \sim Q^{-1} \ln^{2}(1/\alpha_S)$ due to plasma instabilities~\cite{Romatschke:2005pm,Romatschke:2006nk,Romatschke:2005ag,Fukushima:2011nq,Berges:2012cj}. Vacuum fluctuations then add to the classical background field with random phase and comparable magnitude. Consequently, the different gauge field configurations of the ensemble become phase decorrelated by the time $\tau\sim Q^{-1}\ln^2(1/\alpha_S)$ and it is reasonable to conjecture that the system smoothly transitions from a classical field dominated ensemble to a fluctuation dominated regime at this time. 

In suitable gauges, these field configurations correspond to a strongly correlated plasma characterized by a large occupancy $\sim1/\alpha_{S}$ of quasi-particle excitations. Thus while classical-statistical field theory provides a valid description up to times where the typical occupancy falls to order unity, an equivalent description of the dynamics at times $\tau\gtrsim Q^{-1}\ln^2(1/\alpha_S)$ can be achieved within the framework of kinetic theory~\cite{Mueller:2002gd,Jeon:2004dh}.  

In kinetic theory, a number of different scenarios of how thermalization proceeds at weak coupling have been developed~\cite{bib:BMSS,Bodeker:2005nv,bib:KMI,bib:KMII,bib:BGLMV}. 
These allow for different types of solutions of the employed kinetic equations. There is a broad regime of overlap between kinetic theory and non-perturbative classical-statistical field simulations for occupancies that are less than $\alpha_S^{-1}$ but greater than unity. In this work we employ the latter to resolve, within the precision of the numerical solutions, the long-standing question~\cite{Berges:2012ks} of which kinetic scenario is realized in the thermalization process.\\

\textit{Initial conditions.} In the following, we will discuss results of classical-statistical field simulations of a weakly coupled but strongly correlated non-Abelian plasma. We formulate our initial conditions at the proper time $\tau_0 \sim Q^{-1} \ln^{2}(1/\alpha_S)$, where the classical-statistical field configurations are initialized as a superposition of transversely polarized quasi-particle modes\footnote{Note that for simplicity we will consider a system that is homogeneous in the longitudinal rapidity and in the transverse coordinates.}, 
\begin{eqnarray}
\label{eq:quantum2}
A_{\mu}^a(\tau_0,\xvect,\eta)&=&\sum_{\lambda=1,2}\int \frac{\dInt^2 \kvect}{(2\pi)^2}\, \frac{\dInt \nu}{2\pi} \, \sqrt{f(\kvect,\nu,\tau_0)}\, \\
&&\times\left[c_{\lambda,a}^{\kvect\nu}\, \xi^{(\lambda)\kvect\nu+}_{\mu}(\tau_0)\, e^{i \kvect\xvect}\,e^{i\nu\eta}+c.c. \right]\;, \nonumber
\end{eqnarray}
and the corresponding conjugate momenta are given by 
\begin{eqnarray}
\label{eq:quantum2E}
E^{\mu}_a(\tau_0,\xvect,\eta)&=&-\tau_0~g^{\mu\nu}\sum_{\lambda=1,2}\int \frac{\dInt^2 \kvect}{(2\pi)^2}\, \frac{\dInt \nu}{2\pi} \, \sqrt{f(\kvect,\nu,\tau_0)}\,\nonumber \\
&&\times\left[c_{\lambda,a}^{\kvect\nu}\, \dot{\xi}^{(\lambda)\kvect\nu+}_{\nu}(\tau_0)\, e^{i \kvect\xvect}\,e^{i\nu\eta} +c.c.\right]\;. \nonumber \\
\end{eqnarray}
Here $\xi^{(\lambda)\kvect\nu+}_{\mu,a}(\tau)$ denote the (time dependent) transverse polarization vectors of modes with transverse momentum $\kvect$, rapidity wave number $\nu$ and polarization index $\lambda=1,2$ in the non-interacting theory\footnote{The detailed expressions are derived in Appendix A of the companion paper~\cite{Berges:2013fga}.} and $c.c.$ denotes complex conjugation. The classical-statistical ensemble is defined by the distribution of the coefficients $c_{\lambda,a}^{\kvect\nu}$, which satisfy
\begin{eqnarray}
\langle c_{\lambda,a}^{\kvect\nu} c_{\lambda',b}^{*\kvect'\nu'}\rangle=\delta_{\lambda\lambda'}\delta_{ab}~(2\pi)^3~\delta^{(2)}(\kvect-\kvect')\delta(\nu-\nu')\;,   \nonumber \\
\end{eqnarray}
and $\langle c_{\lambda,a}^{\kvect\nu}c_{\lambda',b}^{\kvect'\nu'}\rangle=\langle c_{\lambda,a}^{*\kvect\nu}c_{\lambda',b}^{*\kvect'\nu'}\rangle=0$. These relations are implemented by choosing the coefficients $c_{\lambda,a}^{\kvect\nu}$ as complex Gaussian random numbers in every simulation.

The field configurations in eqns.~(\ref{eq:quantum2}) and (\ref{eq:quantum2E}) describe a weakly coupled plasma of quasi-particle excitations whose properties at the initial time are described by the gluon distribution function $f(p_{T},p_{z},\tau_0)$.  Instead of focusing on one particular realization, we will attempt to capture a large range of different initial conditions by employing a general parametrization
\begin{equation}
\label{eq:quasi-particle}
f(p_{T},p_{z},\tau_0) = \frac{n_0}{8\pi\alpha_S}\, \Theta\!\left( Q - \sqrt{p_T^2+(\xi_0 p_z)^2}\right)\;,
\end{equation}
which describes the overpopulation of gluon modes up to the momentum $Q$.  As long as the dynamics of the non-Abelian plasma at times $\tau \gtrsim  Q^{-1} \ln^{2}(1/\alpha_S)$ is captured by quasi-particle excitations, this parametrization, while perhaps not exhaustive, should capture the physics faithfully. The details of the evolution from earlier times $\tau \sim Q^{-1}$ up to $\tau_0$ are hereby subsumed in particular values of the initial occupancy $n_0$, and in the anisotropy of the initial momentum distribution captured by variations in $\xi_0$. 

The central result of this paper is that, for wide variations of parameters in the above initial conditions, the classical-statistical evolution of the system at late times demonstrates a striking self-similar behavior independent of the initial conditions. Such a self-similar behavior of distributions is characteristic of wave turbulence and, as we shall elaborate further, reflects universal properties of the space-time evolution of the overoccupied non-Abelian plasma. An unanticipated conclusion arises when the classical-statistical simulations are compared to kinetic theory. While the physics of plasma instabilities and free streaming plays an important role in the dynamics at early times, they do not govern the universal turbulent regime. Instead, as we shall discuss, we observe that the first stage of the ``bottom-up" thermalization scenario \cite{bib:BMSS} emerges as a (non-thermal) fixed point of the evolution.\\

\textit{Simulations.} Since the dynamics of the highly occupied plasma is dominated by gluons, we consider a pure Yang-Mills theory in $3+1$ dimensions with longitudinal expansion. We perform real time simulations for the $SU(2)$ gauge group. Previous real time simulations of non-Abelian gauge theories have shown that the $SU(2)$ Yang-Mills dynamics is capturing the relevant dynamics of the $SU(3)$ case~\cite{Krasnitz:2001qu,Berges:2008zt,Ipp:2010uy}. We employ the Kogut-Susskind lattice Hamiltonian in Fock-Schwinger gauge ($A_{\tau}=0$) and solve the classical Hamilton equations of motion on a $N_{T}\times N_{T}\times N_{\eta}$ spatial lattice with periodic boundary conditions.

In the lattice formulation, the continuum gauge fields $A_{\mu}^a(x)$ are represented in terms of the gauge link variables
\begin{eqnarray}
\label{lat:Link}
U_{i}(x)&=&\exp[iga_\bot A_i^a(x+\ihat/2)\Gamma^a]\;, \nonumber \\
U_{\eta}(x)&=&\exp[iga_\eta A_{\eta}^a(x+\hat{\eta}/2)\Gamma^a]\;,
\end{eqnarray}
where $\Gamma^a$ are the generators of the $su(2)$ Lie algebra in the fundamental representation and $g^2 = 4\pi\alpha_s$ denotes the gauge coupling. The symbol $\hat{\mu}=\hat{x}^1,\hat{x}^2,\hat{\eta}$ denotes the neighboring lattice site in the $\mu$ direction, separated by the lattice spacings  $a_{\bot}$ and $a_{\eta}$ in the transverse and longitudinal directions respectively. The conjugate momentum fields $E^{\mu}_{a}(x)$ are represented in terms of the dimensionless electric field variables
\begin{eqnarray}
\label{lat:EField}
\tilde{E}^{i}_{a}(x)=ga_{\bot} E^{i}_{a}(x+\ihat/2+\hat{\tau}/2)\;, \nonumber \\
\tilde{E}^{\eta}_{a}(x)=ga_{\bot}^{2} E^{\eta}_{a}(x+\hat{\eta}/2+\hat{\tau}/2)\;, 
\end{eqnarray}
that are discretized at half-integer time-steps. We then numerically solve the equations of motion derived from the lattice Hamiltonian, with the initial conditions for the real time evolution of the gauge fields and their conjugate momenta provided at the initial time $\tau_0$ by eqs.~(\ref{eq:quantum2}) and (\ref{eq:quantum2E}). The numerical procedure closely follows that employed in previous studies \cite{Romatschke:2005pm,Fukushima:2011nq,Berges:2012cj}. 

Gauge invariant observables are calculated from the standard lattice plaquettes $U_{\mu\nu}(x)=U_{\mu}(x)U^{\dagger}_{\nu}(x+\hat{\mu}-\hat{\nu})U^{\dagger}_{\mu}(x-\hat{\nu})U_{\nu}(x-\hat{\nu})$, with ${\mu,\nu}=(\tau,x_1,x_2,\eta)$ and gauge invariant combinations of the electric field variables such as $E^{2}_{\mu}(x)=\sum_{a} (E_{\mu}^{a}(x))^2$. When we extract gauge dependent quantities we use the residual gauge freedom to perform time independent gauge transformations to impose the generalized Coulomb gauge condition $\partial_iA_i+\tau^{-2}\partial_\eta A_\eta=0$ ($i$=1,2) at each read-out time by use of standard lattice gauge-fixing techniques~\cite{Cucchieri:1995pn}.

We use this procedure to compute the gluon distribution function $f(\pvect,p_z,\tau)$, which describes the occupation number of gluons per momentum mode averaged over spin and color degrees of freedom. Since this quantity has a direct analog in kinetic theory, it is particularly useful to establish a comparison between the different methods.  We extract it using the Fock state projection, 
\begin{eqnarray}
\label{lat:ParticleNumber}
&&f(\pvect,p_z,\tau)=\frac{\tau^2}{N_g V_{\bot}L_{\eta}}\sum_{a=1}^{N_c^2-1}\sum_{\lambda=1,2} \\
&&\Big<\Big|~g^{\mu\nu} \Big[ \Big(\xi^{(\lambda){\pvect\nu+}}_{\mu}(\tau)\Big)^*
\stackrel{\longleftrightarrow}{\partial_{\tau}} 
A^{a}_{\nu}(\tau,\pvect,\nu)
\Big]\Big|^2\Big>_{\text{Coul.~Gauge}}\;, \nonumber
\end{eqnarray}
where the metric reads $g^{\mu\nu} = {\rm diag}(1,-1,-1,-\tau^{-2})$ and the longitudinal momentum at mid-rapidity is identified as $p_z=\nu/\tau$ according to the kinetic term in the field equations. Here the index $\lambda=1,2$ counts the two transverse polarizations, $N_g=2(N_c^2-1)$ denotes the number of transversely polarized gluon degrees of freedom and we write $A\overleftrightarrow{\partial_{\tau}}B=A\partial_{\tau}B-B\partial_{\tau}A$. The gauge field $A^{a}_{\nu}(\tau,\pvect,\nu)$ in eq.~(\ref{lat:ParticleNumber}) is computed from the gauge fixed plaquette variables by inversion of eq.~(\ref{lat:Link}) and a subsequent fast Fourier transformation is performed to obtain the result in momentum space. Similarly, we obtain the time derivative of the gauge field from the fast Fourier transform of the gauge fixed electric field variables in eq.~(\ref{lat:EField}). 

The results presented in this paper are obtained for the initial time chosen as $Q\tau_0=100$ to minimize discretization errors.\footnote{We also explored different choices $Q\tau_0=20 - 1000$. The qualitative behavior observed at late times is very similar to the results for $Q\tau_0=100$.} Since the longitudinal lattice momenta experience a red-shift due to the longitudinal expansion, the maximal longitudinal lattice momentum $p_z^{\max} \sim \pi / a_{\eta} \tau$ decreases in time. Starting the simulations at earlier times thus requires very fine lattice spacing in the longitudinal direction in order to properly resolve the physical scales of the problem. Likewise, the proper resolution of small transverse momenta on the order of the Debye mass $m_D$ requires very large volumes in the transverse direction. Because the physical scales change rapidly with proper time, shifting the initial time $Q\tau_0$ of the simulation to earlier times becomes numerically challenging.

We argued above that our initial conditions are applicable at times $\tau_0 \simeq Q^{-1}\ln^2\alpha_S^{-1}$. This corresponds to extremely small values of $\alpha_S \sim 10^{-5}$ for $Q\tau_0 = 100$. However, in this regard, the following points should be noted:

\noindent
a) 
Within the realm of classical dynamics, all results presented in this paper are independent of the value of the coupling constant $\alpha_S$.

\noindent
b) Since $\alpha_S$ scales out of the classical dynamics, it acts primarily as a parameter to ensure that the occupancy $f \sim 1/\alpha_S$ is larger than the quantum `1/2' -- that this be true for all relevant modes considered, and at all times, is essential for classical-statistical dynamics to provide an accurate description of the system on the time scales of interest~\cite{Moore:2001zf,Aarts:2001yn,Arrizabalaga:2004iw}. 

\noindent
c) As we shall discuss, for the small values of $\alpha_S$ considered, self-similar behavior  is observed for $Q \tau \sim 10^{2}-10^3$. For these values of $\alpha_S$,  we will argue that classical-statistical dynamics is valid up to $Q\tau \sim \alpha_S^{-3/2}\sim 10^8$. As $\alpha_S$ is increased to more realistic values, all the dynamical regimes outlined shrink rapidly. This includes the time scale controlling the growth of plasma instabilities, the lifetime of the classical regime, and possibly the lifetime controlling a preequilibrium quantum regime that may follow. We will return later to a more detailed discussion of this point in the context of our understanding of the thermalization process.

Before we proceed to the discussion of our simulation results, we emphasize for the interested reader that further discussion of details of the numerical implementation, the employed lattice discretization as well as an analysis of possible discretization errors can be found in the companion paper \cite{Berges:2013fga}.\\

\textit{Numerical results.} We first study the time evolution of the bulk anisotropy of the plasma, as described by the components of the gauge invariant stress-energy tensor 
$T^{\mu\nu} = - g^{\nu\alpha} F^{\mu\delta} F_{\alpha\delta} + g^{\mu\nu} F^{\gamma\delta} F_{\gamma\delta}/4$. Specifically, we investigate the volume averages of the transverse and longitudinal pressures $P_T(\tau) = - \left\langle T^x_x(\tau)+T^y_y(\tau) \right\rangle/2$,
$P_L(\tau) = - \left\langle T^\eta_\eta(\tau) \right\rangle$, 
where the brackets denote Monte Carlo averaging with respect to the ensemble of initial field configurations. 

\begin{figure}[t!]						
\includegraphics[width=0.51\textwidth]{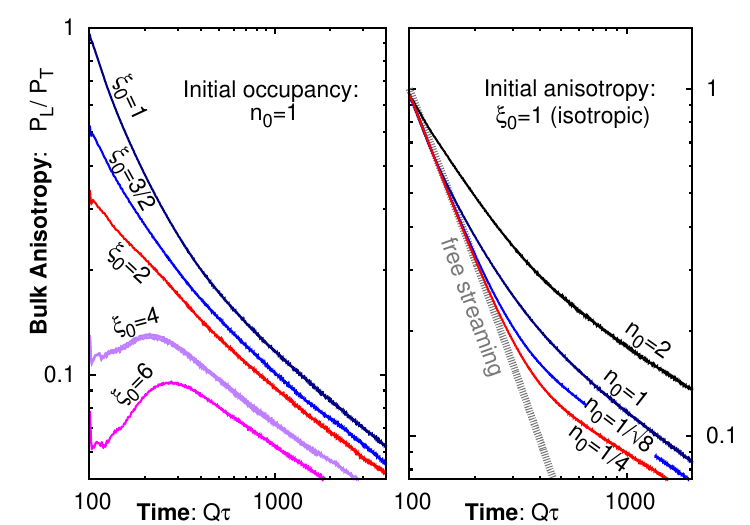}	
\caption{\label{fig:Pressure} Ratio of longitudinal to transverse pressure as a function of time. The left panel shows the result for different initial anisotropies $\xi_0$ and fixed initial occupancy $n_0 = 1$. The right panel shows the same quantity for an initially isotropic system ($\xi_0=1$) and different initial occupancies $n_0$ in comparison to the free streaming (dashed) curve.}	
\end{figure}							

In Fig.~\ref{fig:Pressure} we show the ratio of longitudinal to transverse pressure $P_L/P_T$ as a function of time on a double logarithmic scale. The curves in the left graph are for different initial anisotropies $\xi_0$ and fixed initial occupancy $n_0 = 1$. Starting from an isotropic initial distribution ($\xi_0 = 1$), the system is seen to become more and more anisotropic with time as a consequence of the longitudinal expansion. In the right graph, the free streaming (dashed) curve is shown for comparison, along with results for different initial occupancies $n_0$ for an initially isotropic system.  Indeed, the early-time behavior is governed by free streaming, whereas at later times the anisotropy of the system increases more slowly as a consequence of interactions. In contrast, for strong initial anisotropy such as $\xi_0=4$ and $6$ shown in the left panel, there is a short transient regime where the bulk anisotropy decreases due to plasma instabilities. Nevertheless, the evolution at later times is unaffected and leads to an increasing anisotropy regardless of the degree of initial anisotropy considered. The most striking observations from Fig.~\ref{fig:Pressure} is that, after the transient regime, all curves show very similar scaling with time, irrespective of the choice of initial conditions. 

\begin{figure}[t!]						
\includegraphics[width=0.51\textwidth]{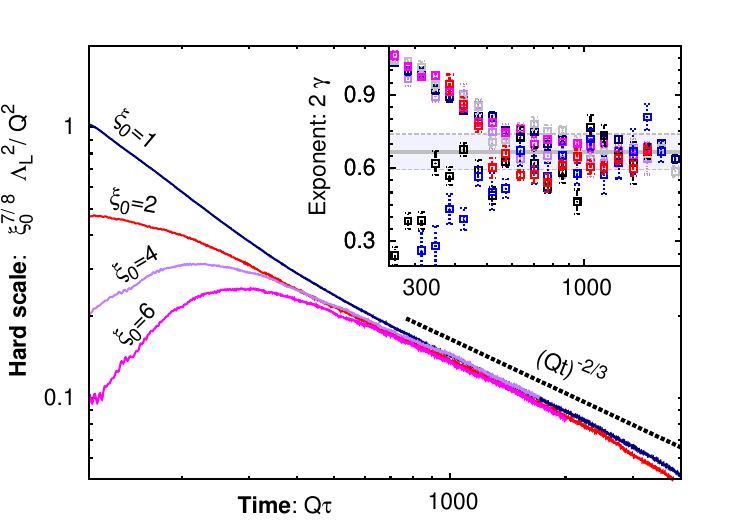}	
\caption{\label{fig:HardScale} Time evolution of the characteristic longitudinal momentum scale for different initial anisotropies $\xi_0$. The inset shows the scaling exponent extracted for different initial conditions and lattice discretizations. The average $2\gamma=0.67\pm0.07$ is indicated by gray lines.}	
\end{figure}							

We will now analyze this behavior in more detail by studying the time evolution of the transverse and longitudinal hard momentum scales $\Lambda_T$ and $\Lambda_L$. These are gauge-invariant quantities computed from transverse and longitudinal projections respectively of the square of the covariant derivative of field strengths divided by the energy density~\cite{Kurkela:2012hp}. The detailed non-perturbative expressions  for the longitudinally expanding case and their lattice implementation are discussed in Ref.~\cite{Berges:2013fga}.  Approximating the field strengths by only their Abelian terms, and relating the gauge fields to the single particle distributions as in eq.~(\ref{lat:ParticleNumber}), one can show that 
\begin{eqnarray}
\label{lat:LambdaApprox}
\Lambda_{T}^2(\tau)&\simeq&\frac{\int \dInt^2\pvect~\dInt p_z~{p_T^2 \over 2}~\omega_p~f(\pvect,p_z,\tau)}{\int \dInt^2\pvect~\dInt p_z~\omega_p~f(\pvect,p_z,\tau)}\;, \nonumber \\
\Lambda_{L}^2(\tau)&\simeq&\frac{\int \dInt^2\pvect~\dInt p_z~p_z^2~\omega_p~f(\pvect,p_z,\tau)}{\int \dInt^2\pvect~d p_z~\omega_p~f(\pvect,p_z,\tau)}\;,
\end{eqnarray}
where $\omega_p\simeq p_T$ is the relativistic quasi-particle energy in the limit $p_T\gg\nu/\tau$. In this limit, it is transparent that $\Lambda_T$ and $\Lambda_L$ characterize the typical momenta of hard excitations in the system. 

In Fig.~\ref{fig:HardScale}, we show the time evolution of the longitudinal hard scale $\Lambda_L^2$ for different initial anisotropies $\xi_0$ and fixed initial occupancy $n_0 = 1$. The different curves are rescaled by an empirical factor of $\xi_0^{7/8}$, to obtain the same normalization in the scaling regime. After the transient regime, for $Q\tau\gtrsim650$,  the typical longitudinal momentum decreases in time and one clearly observes the same power-law behavior irrespective of the initial condition employed.

This scaling behavior of the hard momentum scales can be characterized in terms of the exponents $\gamma$ and $\beta$ as
\begin{equation}
\Lambda_L^2(\tau) \sim (Q \tau)^{-2 \gamma}\, , \quad \Lambda_T^2(\tau) \sim (Q \tau)^{-2 \beta} .
\label{eq:scales}
\end{equation}
The comparison to the dashed curve $\sim (Q \tau)^{-2/3}$ in Fig.~\ref{fig:HardScale} indicates an approximate value of $\gamma \simeq 1/3$. The inset of Fig.~\ref{fig:HardScale} gives the scaling exponent $2\gamma$, extracted from the logarithmic derivative of $\Lambda_L^2$, as a function of time. Results are shown for a set of four different initial conditions in the range \mbox{$\xi_0 = 1$ -- $6$} and $n_0 = 0.25$ -- $2$. To check for a possible further dependence on the lattice discretization, we also display results from the evolution for $\xi_0=n_0=1$ using four different $N_T^2 \times N_\eta$ lattices in the range $N_T = 256$ -- $512$, $N_\eta = 1024$ -- $4096$ with $Q a_T = 0.5$ -- $1$ and $a_\eta = (0.625$ -- $2.5)\cdot 10^{-3}$. By averaging over the data points for $Q \tau \gtrsim 650$ we obtain $2\gamma = 0.67\pm0.07$.

In contrast, in this regime $\Lambda_T$ stays approximately constant in time, which corresponds to $\beta \simeq 0$. Indeed, the exponent extracted from a fit to our data for $\Lambda_T^2$ approaches zero monotonically with 
$|\beta| < 0.06$ for $800\leq Q\tau \leq 2000$.  In general, we observe that the errors are not dominated by discretization errors but the remaining dependence on the initial conditions for the available finite times.\footnote{We also note that, as shown in \cite{Berges:2013fga},  the Debye scale, which is relevant also for the physics of plasma instabilities, is resolved for all simulation times on the large lattices.}

As it is evident from the gauge invariant observables considered, the dynamics becomes independent of the underlying initial conditions at late times. It is a question of general interest whether and how a system effectively loses knowledge of the underlying initial conditions with time. While any thermalization process requires such an effective memory loss when thermal equilibrium is reached, a partial loss of sensitivity to initial conditions may already be observed at earlier stages of the non-equilibrium time evolution. More specifically, a system which is initially far from equilibrium may approach a non-thermal fixed point of the evolution prior to the approach to thermal equilibrium. When this occurs, the subsequent non-equilibrium evolution becomes independent of the details of the underlying initial conditions and is characterized by a self-similar evolution in time~\cite{Micha:2004bv,Berges:2008wm,Berges:2008sr,Nowak:2011sk,Berges:2008mr,Schlichting:2012es,Kurkela:2012hp}. 

In terms of the gluon distribution function a self-similar evolution for the longitudinally expanding system has to fulfill the condition 
\begin{eqnarray}
\label{eq:scalingf}
f(p_T,p_z,\tau)=(Q\tau)^{\alpha}f_S\Big((Q\tau)^\beta p_T,(Q\tau)^\gamma p_z\Big) ,
\end{eqnarray}
where $f_S$ denotes a {\em stationary} distribution independent of time which describes the spectral properties of the non-thermal fixed point. The scaling exponents $\alpha$, $\beta$ and $\gamma$ are pure numbers and characterize the self-similar scaling in time. 

The scaling exponents $\beta$ and $\gamma$ describe the evolution of the characteristic momentum scales as discussed above in the context of eq.~(\ref{eq:scales}). The scaling exponent $\alpha$ describes the overall decrease of the distribution amplitude in time. Even though the latter is not a gauge invariant quantity, the linear combination $\alpha - 3\beta -\gamma$ can be extracted in a gauge invariant fashion from the scaling behavior of the energy density. In the anisotropic scaling limit, one finds
\begin{equation}
\alpha - 3\beta-\gamma = {d\ln \epsilon(\tau)\over d\ln (\tau)} \equiv -\left(1+ {P_L(\tau)\over\epsilon(\tau)}\right) \,.
\end{equation}
One extracts from the behavior of $P_L/\epsilon$ that $\alpha-3\beta-\gamma$ approaches the value $-1$ monotonically from below.\footnote{Since $\epsilon=2P_T+P_L$ for the classical theory this information is already contained in Fig.~(\ref{fig:Pressure}), which shows that $P_L/P_T$ approaches zero monotonically at late times.} We have established in Ref.~\cite{Berges:2013fga} that the residual deviation is $|\alpha(\tau)-3\beta(\tau)-\gamma(\tau)+1| < 0.05$ for $800\leq Q\tau\leq 2000$. 

\begin{figure}[t!]						
 \includegraphics[width=0.47\textwidth]{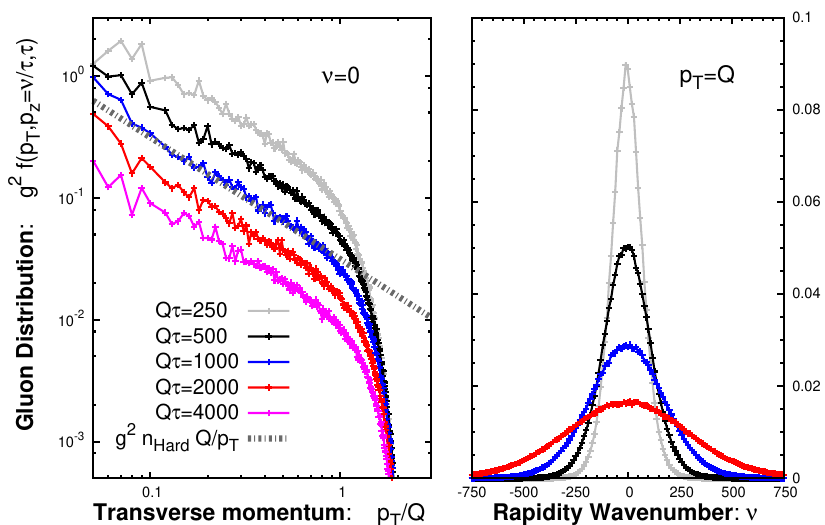}
  \caption{\label{fig:ptSpectra} Gluon distribution function  at different times $Q\tau$ of the evolution. The left graph shows the dependence on transverse momentum $p_T$ for vanishing longitudinal momenta $p_z=0$ at different times. The right graph shows the same distribution as a function of rapidity wave number $\nu=\tau p_z$ for modes with transverse momenta $p_T\simeq Q$.}
\end{figure}
We will now investigate the emergence of self-similarity in our simulations and directly study the time evolution of the gluon distribution function $f(p_T,p_z,\tau)$ extracted from the Coulomb gauge fixed fields as in eq.~(\ref{lat:ParticleNumber}). Fig.~\ref{fig:ptSpectra} shows the gluon distribution function at different times $Q\tau$ of the evolution. The left graph displays the distribution as a function of transverse momentum for modes with vanishing longitudinal momenta $p_z=0$ at different times. The spectrum shows a $1/p_T$ power-law over a large range of transverse momenta \mbox{$p_T\lesssim\Lambda_T$} with a rapid fall-off for  large momenta $p_T\gtrsim\Lambda_T$. It is interesting to note that a two-dimensional Bose-Einstein distribution would exhibit the same $1/p_T$ behavior in the classical regime of low momenta $p_T < T$, where $T$ is the temperature. The spectrum exhibits a self-similar evolution, where the position of hard momentum scale $\Lambda_T$, demarcating the high momentum fall-off, is seen to remain approximately constant in time in accordance with our previous discussion of $\beta\simeq0$. The over-all amplitude of the distribution $n_{\rm Hard}(t)=f(p_T=Q,p_z=0,\tau)$ is seen to decrease in time. 

The right graph in Fig.~\ref{fig:ptSpectra}, shows the gluon distribution for modes with $p_T=Q$ as a function of rapidity wave number $\nu=\tau p_z$. The overall shape is well described by a Gaussian distribution, with a decreasing amplitude and increasing width in time. We emphasize that this behavior is very different from free-streaming, where the distribution function $f(p_T,p_z=\nu/\tau,\tau)$ would be time independent when plotted as a function of transverse momentum $p_T$ and rapidity wave number $\nu$. Instead of such a constant behavior in the non-interacting theory, one clearly observes from Fig.~\ref{fig:ptSpectra} that interactions lead to a redistribution of particles from smaller to higher rapidity wave numbers.

\begin{figure}[t!]						
 \includegraphics[width=0.49\textwidth]{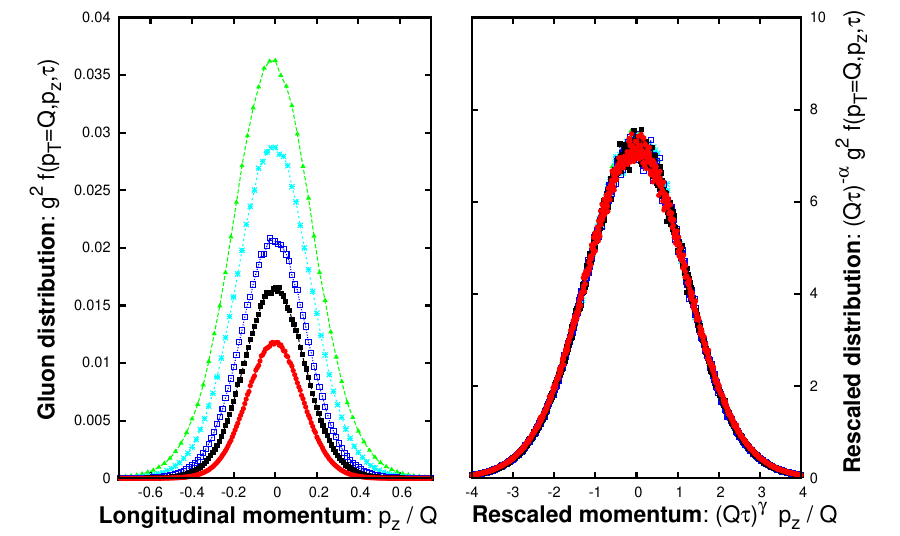}
  \caption{\label{fig:self-similar} (\textbf{left}) Gluon distribution as a function of longitudinal momentum $p_z$ at different times  $Q\tau=750,1000,1500,2000,3000$ (top to bottom)  of the evolution. (\textbf{right}) The same data is shown after rescaling of the distribution function and the longitudinal momentum. The rescaled data for different times all collapses onto a single curve.}
\end{figure}
The self-similarity of this process can be demonstrated by investigating the scaling properties of the rescaled moments of the single particle distribution. In the right panel of Fig.~\ref{fig:self-similar} we show the rescaled gluon distribution as a function of rescaled longitudinal momenta $(Q\tau)^\gamma p_z$ for modes with transverse momenta $p_T\simeq Q$. The left panel of Fig.~\ref{fig:self-similar} shows the original data for comparison. Once the distribution satisfies the self-similarity condition in eq.~(\ref{eq:scalingf}), this simple rescaling should account for the dynamical evolution between the different times shown in Fig.~\ref{fig:self-similar}. Indeed, the rescaled data for different times in the range from $Q\tau=750$ to $4000$ is seen to collapse onto a single curve to high accuracy, which is a striking manifestation of the self-similarity of the evolution.\footnote{A detailed discussion of the scaling exponents inferred from this self-similar behavior, taking into account systematic errors, is provided in our companion paper~\cite{Berges:2013fga}.}

\textit{Kinetic theory analysis.} The self-similar scaling behavior extracted from both gauge invariant and gauge fixed observables in the classical-statistical field simulations of the expanding non-Abelian plasma finds a simple a posteriori explanation in the context of wave turbulence following a previous analysis \cite{Micha:2004bv} 
in the context of scalar field theories. This turbulence analysis is performed in the framework of a kinetic equation 
\begin{eqnarray}
\label{eq:Boltzmann}
\left[\partial_{\tau}-\frac{p_z}{\tau}\partial_{p_z}\right]f(p_T,p_z,\tau)=C[p_T,p_z,\tau;f] \;,
\end{eqnarray}
for the single particle distribution $f(p_T,p_z,\tau)$ with a generic collision term $C[p_T,p_z,\tau;f]$ for $n\leftrightarrow m$ scattering processes. 
For the self-similar distribution in eq.~(\ref{eq:scalingf}),
the scaling behavior of the collision integral 
\begin{equation}
C[p_T,p_z,\tau;f] = (Q \tau)^{\mu}\, C[(Q \tau)^{\beta}p_T,(Q \tau)^{\gamma}p_z;f_S] \, ,
\end{equation}
in this analysis is described in terms of the exponent $\mu=\mu(\alpha,\beta,\gamma)$, whose precise form depends on the underlying interaction~\cite{Micha:2004bv}. Substituting this form into the Boltzmann equation (\ref{eq:Boltzmann}) leads to the time-independent condition
\begin{eqnarray}
\label{eq:attractor}
&& \alpha f_S(p_T,p_z)
+\beta p_T \partial_{p_T} f_S(p_T,p_z) \nonumber\\
&& +  \left(\gamma-1\right) p_z \partial_{p_z} f_S(p_T,p_z)
= \OneOverQ C[p_T,p_z;f_S]\;. 
\end{eqnarray}
The non-thermal attractor solution observed in our lattice simulations corresponds to a nontrivial solution of this equation when phrased in terms of a kinetic description.

Simultaneously, one obtains the scaling relation
\begin{eqnarray} 
\label{eq:srel}
\alpha-1=\mu(\alpha,\beta,\gamma)\;.
\end{eqnarray}
which constraints the self-similar evolution along the attractor. To obtain specific values of the scaling exponents $\alpha$, $\beta$ and $\gamma$ that agree with our classical-statistical lattice results, one needs to determine the dominant type of scattering processes in the kinetic theory analysis. At this juncture, we shall follow the bottom-up thermalization scenario~\cite{bib:BMSS} which, as we will shortly show, provides the best agreement with our classical-statistical lattice results. In the high occupancy classical regime of the bottom-up scenario, the interaction of hard quasi-particle excitations is dominated by elastic scattering with small momentum transfer. Inelastic interactions primarily affect the soft sector as long as the occupancies of the hard excitations are large $(n_{\text{Hard}}\gg1)$ and plasma instabilities were not considered in that work. The effect of elastic collisions is to broaden the longitudinal momentum distribution by multiple incoherent small-angle scatterings. The collision integral can then be approximated to be of the Fokker-Planck type,
\begin{eqnarray}
C^{({\rm elast})}[p_T,p_z;f]=\hat{q}~\partial_{p_z}^2 f(p_T,p_z,\tau),
\end{eqnarray}
where the momentum diffusion parameter is approximated as 
\begin{eqnarray}
\hat{q}\sim \alpha_S^2 N_c^2 \int {d^2p_T\over (2\pi)^2}\int {dp_z \over 2\pi}~f^2(p_T,p_z,\tau)
\end{eqnarray}
for high occupancies. Although this approximation may not capture all details, it is expected to describe the relevant physics necessary to determine the scaling exponents. 

While the original work by Baier,~Mueller,~Schiff and Son~\cite{bib:BMSS} (BMSS) determines the basic properties of the kinetic evolution from self-consistency arguments, the self-similar behavior observed from numerical simulations indicates that the framework of turbulent thermalization~\cite{Micha:2004bv} can be applied. We continue this analysis by plugging the self-similar distribution (\ref{eq:scalingf}) into $C^{({\rm elast})}[p_T,p_z;f]$ to extract the scaling behavior $\mu = 3\alpha-2\beta+\gamma$. The scaling relation in eq.~(\ref{eq:srel}), then reads $2\alpha-2\beta+\gamma+1 = 0$. Since elastic scattering processes are particle number conserving, a further scaling relation is obtained from integrating the distribution function over $p_T$ and rapidity wave numbers $\nu = p_z \tau $. By use of the scaling form (\ref{eq:scalingf}), particle number conservation leads to the scaling relation $\alpha-2\beta-\gamma+1=0$. Similarly, approximating the mode energy of hard excitations as $\omega_p \simeq p_T$ in the anisotropic scaling limit,  energy conservation yields the final scaling condition $\alpha-3\beta-\gamma+1=0$.

Remarkably, the above scaling relations are independent of many of the details of the underlying field theory such as the number of colors, the coupling constant as well as the initial conditions. Instead, they only depend on the dominant type of kinetic interactions (such as $2\leftrightarrow2$ or $2\leftrightarrow3$ scattering processes), the conserved quantities of the system and the number of dimensions. More specifically, the dynamics of small-angle elastic scattering, along with the conservation laws of quasi-particle number and energy provide the three equations to determine the scaling exponents. These are straightforwardly extracted to be 
\begin{eqnarray}
\alpha=-2/3\;, \qquad \beta=0\;, \qquad  \gamma=1/3 \;,
\end{eqnarray}
in good agreement with those extracted from our lattice simulations of the temporal evolution of gauge invariant observables.

\begin{figure}[t!]						
\vspace*{-3.5ex}
\includegraphics[width=0.48\textwidth]{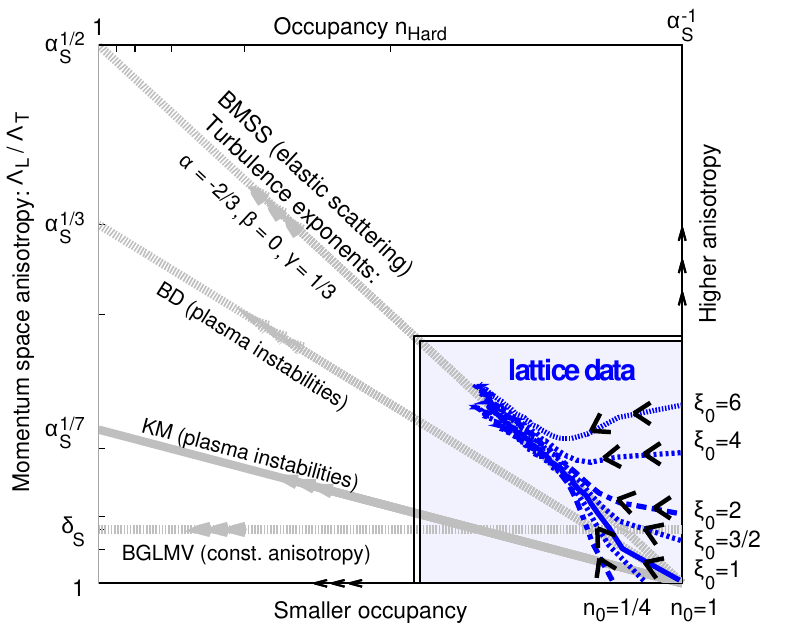}	
\caption{\label{fig:Cartoon} Evolution in the occupancy--anisotropy plane. Indicated are the attractor solutions proposed in (BMSS) \cite{bib:BMSS}, (BD) \cite{Bodeker:2005nv}, (KM) \cite{bib:KMII} and (BGLMV) \cite{bib:BGLMV}, along with the simulations results for different initial conditions shown in blue.}							
\end{figure}							

The close agreement of the lattice simulations with the bottom-up scenario appears surprising at first. While in the latter, it is the Debye scale that provides the scale for multiple incoherent elastic scatterings and the consequent broadening of the longitudinal momentum, the one loop self-energy for anisotropic momentum distributions could lead to plasma instabilities even at times $\tau\gtrsim Q^{-1} \log^{2}(\alpha_S^{-1})$. The impact of plasma instabilities on the first stage of the bottom-up scenario has been considered in ~\cite{Bodeker:2005nv} (BD). In this scenario, plasma instabilities create an overpopulation of the unstable soft modes $f(p\sim m_D)\sim 1/\alpha_S$, such that the interaction of hard excitations with the highly populated soft modes becomes the dominant process. This process leads to a more efficient momentum broadening in the longitudinal direction and changes the evolution of the characteristic momentum scales and occupancies. Similar considerations, albeit including a different range of highly occupied unstable modes\footnote{The range of highly occupied unstable modes in this scenario is determined within the hard-loop framework in Ref.~\cite{bib:KMI} and parametrically given by modes with momenta $p_T\lesssim m_D$ and $p_z\lesssim m_D \Lambda_T/\Lambda_L$.}, lead to the detailed weak coupling scenario in \cite{bib:KMII} (KM). In this scenario, plasma instabilities play a significant role for the entire thermalization process in the classical regime and beyond. Yet another scenario of how highly occupied expanding non-Abelian fields proceed toward thermalization was proposed in \cite{bib:BGLMV}. In this scenario, it is conjectured that the combination of high occupancy and elastic scattering can generate a transient Bose-Einstein condensate. The evolution of this condensate together with elastically scattering quasi-particle excitations is argued to generate an attractor with fixed $P_L/P_T$ anisotropy parameter $\delta_s$. 

While all of these effects can in principle be realized and have interesting consequences for the subsequent space-time evolution of the strongly correlated plasma, the infrared physics of momenta around the Debye scale is crucial in all these scenarios. The properties of this highly non-linear non-Abelian dynamics can be resolved conclusively through non-perturbative numerical simulations, such as those performed here.

A compact summary of our results in comparison with the different weak coupling thermalization scenarios is shown in Fig.~\ref{fig:Cartoon}, describing the space-time evolution in the occupancy--anisotropy plane. The horizontal axis shows the occupancy $n_{\text{Hard}}$ and the vertical axis the momentum-space anisotropy in terms of the typical longitudinal and transverse momenta $\Lambda_{T,L}$. The gray lines indicate the attractor solutions of the different thermalization scenarios, while the blue lines show our simulation results for different initial conditions. One immediately observes the attractor property, which appears to be in good agreement with the analytical discussion of the BMSS kinetic equation in the high-occupancy regime~\cite{Berges:2013fga}.

As noted previously, similar attractor solutions were discovered in relativistic scalar theories that purport to describe the highly occupied post-inflationary thermalization phase of the early universe. Scaling analyses of kinetic equations, identical to those in studies of weak wave turbulence, demonstrated that the scaling exponents characterizing the attractor could be classified on very general grounds of dimensionality, conservation laws and boundary conditions for the evolution~\cite{Micha:2004bv,Berges:2008wm}.  Systems on diverse energy scales, ranging from the discussed early-universe inflaton dynamics of relativistic scalar fields to table-top experiments with cold atoms, have universal scaling exponents that characterize self-similar attractor solutions~\cite{Berges:2008wm,Berges:2008sr,Nowak:2011sk}.
The observation here of such a self-similar scaling solution of the expanding non-Abelian plasma is a powerful indication for universal behavior far from equilibrium. In light of this discussion, one can conclude that the bottom-up scenario correctly captures the universal properties of the turbulent thermalization process observed in our simulations. However, it is also conceivable that there are processes besides small angle elastic scattering which lead to the same scaling behavior.\\ 

\begin{figure}[t!]	 
\vspace*{-3.5ex}
\includegraphics[width=0.48\textwidth]{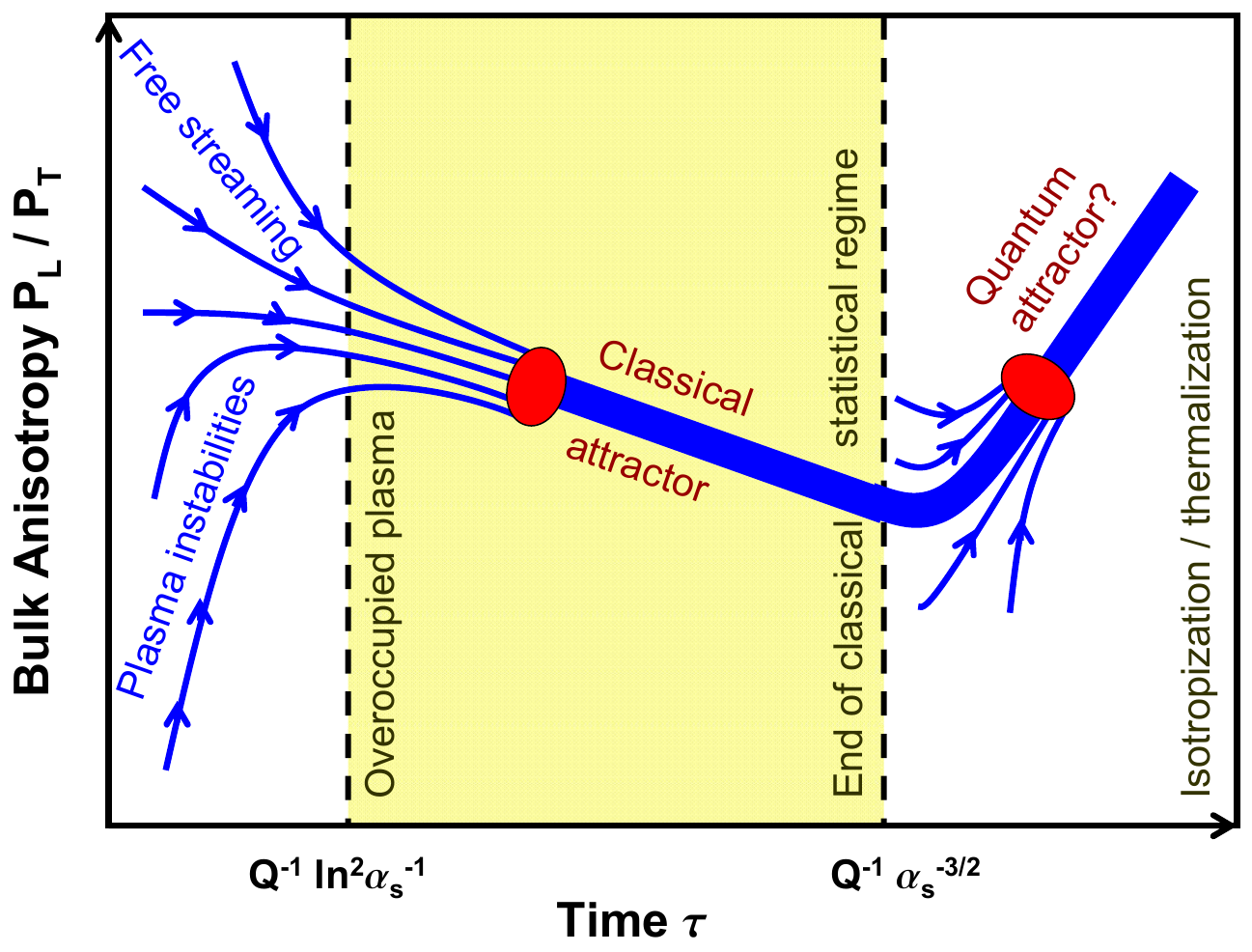}	
\caption{\label{fig:Summary} Schematic illustration of the thermalization process for the example of the bulk anisotropy. The time evolution in the shaded yellow regime pictures the results of this work, with the red ellipse symbolizing the observed non-thermal fixed point. The curves beyond the classical regime reflect the evolution in the bottom-up scenario \cite{bib:BMSS}, with the possibility of a second turbulent attractor in the quantum regime~\cite{Blaizot:2013hx}.}	
\end{figure}	 

\textit{Summary \& Conclusions.} In this paper, we discussed a first principles study of the dynamics of a highly occupied non-Abelian plasma using large scale numerical simulations. The discovery of a universal attractor in a temporal regime where the occupancy of the plasma is still large, and the emergent self-similar behavior provide unprecedented insights into the dynamics of the longitudinally expanding plasma. We have also established for the first time a link between non-perturbative classical-statistical field simulations and weak coupling thermalization scenarios formulated in kinetic theory. Somewhat surprisingly, we find that, within their range of applicability, the simulation results show a clear preference for the bottom-up thermalization scenario \cite{bib:BMSS} where the quasi-particle dynamics is governed predominantly by elastic scattering. 

The emergent picture of the thermalization process at weak coupling is illustrated schematically in Fig.~\ref{fig:Summary} at the example of the bulk anisotropy of the plasma. Following a transient regime dominated by plasma instabilities and free streaming, an effective memory loss is observed already at rather early times of the evolution and leads to a turbulent attractor solution for different initial conditions. Our results also suggest that the plasma, albeit strongly interacting at all times, becomes increasingly anisotropic in the classical-statistical regime of high occupancies. Consequently, the system is still far from equilibrium at the transition from the classical regime to the quantum regime\footnote{Note that, as previously observed, the system exhibits a Bose-like $1/p_T$ distribution already well before the end of the classical regime.}. Using the observed scaling behavior, this is expected to occur on a time scale $\tau\sim Q^{-1} \alpha_S^{-3/2}$. Since classical-statistical simulations are no longer applicable beyond this point, this leaves several open questions that need to be addressed in the future to deepen our understanding of the thermalization process.

Most importantly, the questions of how and on what time scales the system may isotropize and approach thermal equilibrium in the quantum regime are beyond the validity of classical-statistical simulations. Kinetic approaches are however valid, and a more detailed description of isotropization and thermalization can be based on such frameworks. In the bottom-up scenario, $2\leftrightarrow3$ scattering processes become increasingly important in the quantum region and lead to isotropization and thermalization on a time scale $\tau\sim Q^{-1}\alpha_S^{-13/5}$, which corresponds to $Q\tau \sim 20$ for $\alpha_S\simeq0.3$ ($\tau\sim2$ fm/c for $Q=2$ GeV). Interestingly, it has been shown recently that such radiative $2\leftrightarrow3$ processes also generate a self-similar cascade~\cite{Blaizot:2013hx}. If so, turbulent dynamics will be responsible for the entire non-equilibrium evolution of the plasma, from far off-equilibrium to thermalization. This possible scenario is also sketched in Fig.~\ref{fig:Summary}.

Of course, it would also be interesting to extend the use of classical-statistical simulations to larger couplings and earlier initialization times to achieve a direct description of heavy-ion collisions starting at earliest times. However, this is not unambiguous given the conceptual limitations of the classical-statistical approach. Indeed the primary objective of working at very weak coupling is to cleanly separate the dynamics of classical modes from quantum evolution effects. While classical-statistical field theory provides a solid description for highly occupied modes $f \gg 1$, quantum evolution effects are extremely important for modes with $f\lesssim 1$. While initially the single particle occupancies $f$ are $\sim 1/\alpha_S$ shortly after $Q\tau_0\sim \ln^2 \alpha_S^{-1}$, the occupancy of typical modes rapidly decreases with time. Unless the single particle occupancies $f$ are initially very large (corresponding to very small values of $\alpha_S$), quantum effects quickly become important. The extrapolation of the classical-statistical field theory to larger couplings must thus be done with great care to ensure robust results. While larger couplings also imply the relevance of earlier time scales $Q\tau_0$, numerical simulations at these early times may require lattices that are significantly larger than those in the present study in order to resolve all relevant modes.\\

\textit{Acknowledgments}. We would like to thank 
J.-P.~Blaizot, K.~Dusling, T.~Epelbaum, F.~Gelis, A.~Kurkela, J.~Liao, L.~McLerran, G.~Moore and D.~Sexty for very valuable discussions. R.V.\ is supported by the US Department of Energy under DOE Contract No.\ DE-AC02-98CH10886. We acknowledge support by the DFG and the MWFK Baden-W\"{u}rttemberg (bwGRiD cluster).

\end{document}